\documentclass[a4paper]{PoS}

\usepackage{epstopdf}
\usepackage{amsmath}
\usepackage{physics}
\usepackage{array}
\usepackage{cite}
\usepackage{placeins}
\usepackage{subfig}
\usepackage{xcolor}

\title{Control of SU(3) symmetry breaking effects in calculations of B meson decay constant}

\ShortTitle{Control of SU(3) breaking effects in calculations of $f_B$}

\author{\speaker{Sophie Hollitt},$^a$
	R.~Horsley,$^b$
	P.~D.~Jackson,$^a$
	Y.~Nakamura,$^c$
	H.~Perlt,$^d$
	P.~E.~L.~Rakow,$^e$
	G.~Schierholz,$^f$
	A.~Schiller,$^d$
	H.~St\"uben,$^g$
	R.~D.~Young$^a$
	and J.~M.~Zanotti$^a$\\
	\llap{$^a$}CSSM/CoEPP, Department of Physics, University of Adelaide, Adelaide SA 5005, Australia\\
	\llap{$^b$}School of Physics and Astronomy, University of Edinburgh, 
	Edinburgh EH9 3FD, United Kingdom\\
	\llap{$^c$}RIKEN Advanced Institute from Computation Science, Kobe, Hyogo 650-0047, Japan\\
	\llap{$^d$}Institut f\"ur Theoretische Physik, Universit\"at Leipzig, 04103 Leipzig, Germany\\
	\llap{$^e$}Theoretical Physics Division, Department of Mathematical Physics,
	University of Liverpool, Liverpool L69 3BX, United Kingdom\\
	\llap{$^f$}Deutsches Elektronen-Synchrotron DESY, 23603 Hamburg, Germany\\
	\llap{$^g$}RRZ, Univeristy of Hamburg, 20146 Hamburg, Germany\\

        E-mail: \email{sophie.hollitt@adelaide.edu.au}}

\abstract{Early $B$-physics experiments have left us with a number of puzzles in heavy flavour physics. New lattice calculations (with a greater understanding of QCD effects in the Standard Model) will be needed to support the increase in experimental precision to be achieved by upcoming experiments such as Belle II. We extend the CSSM/UKQCD/QCDSF studies of SU(3) flavour breaking effects by presenting new results for the decay constants $f_B$ and $f_{B_s}$.
}

\FullConference{The 36th Annual International Symposium on Lattice Field Theory - LATTICE2018\\
		22-28 July, 2018\\
		Michigan State University, East Lansing, Michigan, USA.}

\begin{document}

\section{Introduction}
Recent $B$ physics experiments have left us with a number of heavy flavour physics puzzles\cite{B-anomalies}, and as Belle II's physics run approaches, a more precise understanding of the QCD contribution to the Standard Model determination of heavy flavour observables is needed to help isolate possible new physics by reducing theory errors. $B$ meson decay constants in particular are used in the determination of multiple CKM matrix elements: $\abs{V_{td}}$ and $\abs{V_{ts}}$ from $B^0 \overline{B^0}$ and $B_s^0 \overline{B_s^0}$ oscillations, and $\abs{V_{ub}}$ and $\abs{V_{cb}}$ from leptonic decays of $B$ and $B_c$ mesons respectively. Combined with precise lattice calculations of $f_B$, measurements of the $B \rightarrow \tau \nu$ branching ratio at Belle II can be used as an independent measurement of $\abs{V_{ub}}$ and give further insight into the existing discrepancy between $\abs{V_{ub}}$ measured from inclusive and exclusive decays.

As $f_B$ is often calculated on the lattice via the ratio $f_{B_s}$/$f_B$\cite{FLAG2016}, it is important to understand and control SU(3) breaking effects in the light and strange quarks, and study how these affect extrapolations of $f_{B_s}$/$f_B$. In this work, we compute $f_B$ and $f_{B_s}$ using a set of gauge field configurations that break SU(3) flavour in a controlled way, keeping the average of the lighter quark masses held fixed at the physical value.

\section{Simulation Details}

\subsection{SU(3) breaking and quark actions}
\begin{figure}[htbp]
	\centering
	\includegraphics[width=0.9\textwidth]{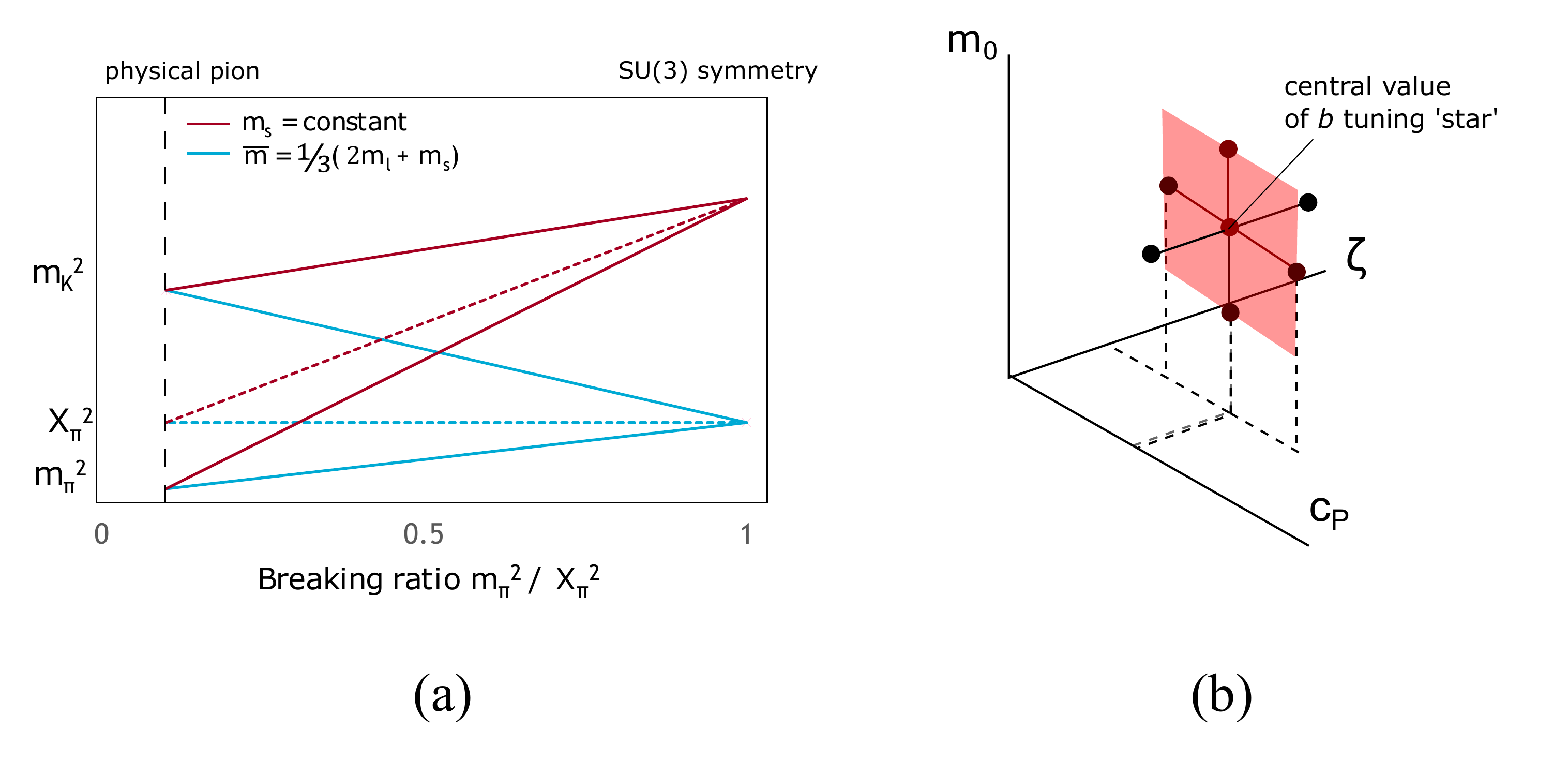}
	\caption{\label{breakingDiagram}(a) Diagram of the evolution of $m_\pi^2$ and $m_K^2$ for constant strange quark mass $m_s$ (red) or constant average quark mass (blue). The dashed lines show the evolution of the flavour singlet $X_\pi^2 = \frac{1}{3}(2m_K^2 + m_\pi^2)$ (b) Diagram of the tuning `star' shape, which has one central value of ($m_0$,$c_P$,$\zeta$) and then additional values at $\pm \Delta m$, $\pm \Delta c_P$, $\pm \Delta \zeta$}.
\end{figure}

We use multiple ensembles of gauge field configurations with 2+1 flavours of non-perturbatively $\mathcal{O}(a)$ improved Wilson fermions. When extrapolating to the physical point using multiple lattice ensembles with different quark masses, it is common practice to choose the strange quark mass $m_s$ to be held (approximately) fixed at its physical value. We instead follow the QCDSF process for choosing the masses of light and strange quarks in a 2+1 flavour formalism\cite{QCDSFstyle}, where the value of $\overline{m} = \frac{1}{3}(2m_l + m_s)$ is kept constant to control symmetry breaking. In this approach, all flavour-singlet quantities are only affected by SU(3)-flavour breaking effects at $\mathcal{O}((\delta m)^2)$, and have been shown to stay approximately constant from the SU(3) symmetric point to the physical point\cite{QCDSFstyle}. A diagram showing this behaviour compared to a standard approach is given in the left plot of Figure \ref{breakingDiagram}. The $x$-axis is scaled with the flavour-singlet combination of light pseudoscalar meson masses, $X_\pi^2 = \frac{1}{3}(2m_K^2 + m_\pi^2)$.

In the specific case of $B$-mesons, we also expect flavour-singlet combinations of $B$ meson properties to be approximately constant along this quark mass trajectory. We can  thus use properties of the physical $B$ flavour singlet as an appropriate target in tuning our $B$-mesons on the lattice. We label this $B$ flavour singlet meson as $X_B = 1/3~(2B_l + B_s)$ and then consider its mass ($M_{X_B}$) or decay constant ($f_{X_B}$) with an appropriate substitution.

We generate bottom quarks using a variant of the `Fermilab action' or `RHQ action'\cite{1997,RHQaction}. This anisotropic clover-improved action has the form\cite{NPtuning}
\begin{equation*}
	S_{lat} = a^4 \sum\limits_{x,x'} \overline{\psi} (x') \left( \vphantom{\sum\limits_{x}} m_0 + \gamma_0 D_0 + \zeta \vec{\gamma} \cdot \vec{D} - \frac{a}{2}(D^0)^2  - \frac{a}{2}\zeta (\vec{D})^2 + \sum\limits_{\mu,\nu} \frac{ia}{4} c_P \sigma_{\mu\nu} F_{\mu\nu} \right) _{x,x'} \psi (x)
\end{equation*}
where $m_0$, $c_P$, and $\zeta$ are tuned as three free parameters. The `best' $B$ meson is selected by tuning the free parameters until the masses and hyperfine splitting of our calculated $X_B$ and $X_{B^*}$ mesons match the properties of the physical $X_B$ and $X_{B*}$.

In practice, uncertainties on measured masses and splittings also result in uncertainty in the values of $m_0$, $c_P$, and $\zeta$ corresponding to the `best' tuned $B$ meson. We choose to generate multiple $b$-quarks per lattice ensemble in a `tuning star' shape (see right-hand plot in Figure \ref{breakingDiagram}) and interpolate to the `best' $B$, rather than generating only one `best' $b$-quark per ensemble. Since we tune using SU(3) flavour singlets, we can employ the same set of seven $b$-quarks for each ensemble with the same lattice spacing and volume along the line of constant $\overline{m}$.

\subsection{Plateau fitting for multiple $\mathbf{b}$ quarks}
$B$-meson properties are calculated using fits to correlators and ratios of correlators. To maintain consistency, we choose to use the same fit window for each correlator across all 7 $b$-quark candidates on a fixed lattice ensemble. Selecting the best fit is assisted by using the correlated $\chi^2/\text{d.o.f}$ for the fit on each $B$ meson correlator.

\subsection{Lattice spacings and volumes}
A variety of lattice spacings and lattice volumes are used in this work. Some details of the QCDSF gauge field ensembles are presented in Table \ref{datatable}. The interpolated `best' values for $m_0$, $c_P$, and $\zeta$ on each ensemble are not shown here due to space constraints, but will be presented in future studies on the systematic uncertainties. For all results, the source locations for the calculated mesons are randomised to reduce correlations between neighbouring configurations in the ensemble.

\begin{table}[h]
\begin{tabular}{cccccccl}
	
	$\beta$ & $a$ (fm) & Lattice volume & $\kappa_{\text{light}}$ & $\kappa_{\text{strange}}$ &  $m_\pi$ (MeV) &  $m_K$ (MeV) &            \\\hline
	
	\multicolumn{ 1}{c}{5.4} & \multicolumn{ 1}{c}{0.082} & \multicolumn{ 1}{c}{$24^3\times48$} &    0.11993 &    0.11993 &        413 &        413 &            \\
	
	\multicolumn{ 1}{c}{} & \multicolumn{ 1}{c}{} & \multicolumn{ 1}{c}{} &   0.120048 &   0.119695 &        325 &        448 &            \\\cline{3-8}
	
	\multicolumn{ 1}{c}{} & \multicolumn{ 1}{c}{} & \multicolumn{ 1}{c}{$32^3\times64$} &    0.11993 &    0.11993 &        408 &        408 &            \\
	
	\multicolumn{ 1}{c}{} & \multicolumn{ 1}{c}{} & \multicolumn{ 1}{c}{} &   0.119989 &   0.119812 &        366 &        424 &          $\dagger$ \\
	
	\multicolumn{ 1}{c}{} & \multicolumn{ 1}{c}{} & \multicolumn{ 1}{c}{} &   0.120084 &   0.119623 &        290 &        450 &          $\dagger$ \\\hline
	
	\multicolumn{ 1}{c}{5.5} & \multicolumn{ 1}{c}{0.074} & \multicolumn{ 1}{c}{$32^3\times64$} &     0.1209 &     0.1209 &        468 &        468 &          * \\
	
	\multicolumn{ 1}{c}{} & \multicolumn{ 1}{c}{} & \multicolumn{ 1}{c}{} &    0.12104 &    0.12062 &        357 &        505 &          * \\
	
	\multicolumn{ 1}{c}{} & \multicolumn{ 1}{c}{} & \multicolumn{ 1}{c}{} &   0.121095 &   0.120512 &        315 &        526 &          * \\
	
		\multicolumn{ 1}{c}{} & \multicolumn{ 1}{c}{} & \multicolumn{ 1}{c}{} &   0.121145 &   0.120413 &        258 &        537 & *           \\\cline{3-8}
	
	\multicolumn{ 1}{c}{} & \multicolumn{ 1}{c}{} & \multicolumn{ 1}{c}{$32^3\times64$} &    0.12095 &    0.12095 &        403 &        403 &            \\
	
	\multicolumn{ 1}{c}{} & \multicolumn{ 1}{c}{} & \multicolumn{ 1}{c}{} &    0.12104 &    0.12077 &        331 &        435 &            \\
	
	\multicolumn{ 1}{c}{} & \multicolumn{ 1}{c}{} & \multicolumn{ 1}{c}{} &   0.121099 &   0.120653 &        270 &        454 &            \\\cline{3-8}
	
	\multicolumn{ 1}{c}{} & \multicolumn{ 1}{c}{} &    $48^3\times96$ &   0.121166 &   0.120371 &        226 &        539 &  *          \\\hline
	
	\multicolumn{ 1}{c}{5.65} & \multicolumn{ 1}{c}{0.068} & \multicolumn{ 1}{c}{$32^3\times64$} &   0.122005 &   0.122005 &        421 &        421 &            \\
	
	\multicolumn{ 1}{c}{} & \multicolumn{ 1}{c}{} & \multicolumn{ 1}{c}{} &   0.122078 &   0.121859 &        361 &        448 &            \\
	
	\multicolumn{ 1}{c}{} & \multicolumn{ 1}{c}{} & \multicolumn{ 1}{c}{} &    0.12213 &   0.121756 &        310 &        463 &            \\\cline{3-8}
	
	\multicolumn{ 1}{c}{} & \multicolumn{ 1}{c}{} & \multicolumn{ 1}{c}{$48^3\times96$} &   0.122005 &   0.122005 &        412 &        412 &            \\
	
	\multicolumn{ 1}{c}{} & \multicolumn{ 1}{c}{} & \multicolumn{ 1}{c}{} &   0.122078 &   0.121859 &        355 &        441 &            \\
	
	\multicolumn{ 1}{c}{} & \multicolumn{ 1}{c}{} & \multicolumn{ 1}{c}{} &    0.12213 &   0.121756 &        302 &        457 &            \\
	
	\multicolumn{ 1}{c}{} & \multicolumn{ 1}{c}{} & \multicolumn{ 1}{c}{} &   0.122167 &   0.121682 &        265 &        474 &            \\\cline{3-8}
	
	\multicolumn{ 1}{c}{} & \multicolumn{ 1}{c}{} &    $64^3\times96$ &   0.122227 &   0.121563 &        155 &        480 &          $\dagger$ \\\hline
	
	\multicolumn{ 1}{c}{5.8} & \multicolumn{ 1}{c}{0.059} & \multicolumn{ 1}{c}{$48^3\times96$} &    0.12281 &    0.12281 &        427 &        427 &            \\
	
	\multicolumn{ 1}{c}{} & \multicolumn{ 1}{c}{} & \multicolumn{ 1}{c}{} &    0.12288 &    0.12267 &        357 &        456 &            \\
	
	\multicolumn{ 1}{c}{} & \multicolumn{ 1}{c}{} & \multicolumn{ 1}{c}{} &    0.12294 &   0.122551 &        280 &        477 &            \\
	
\end{tabular}
\caption{Table of lattice ensembles used in this work. * indicates ensembles with a different value of $\overline{m}$, further from the physical $\overline{m}$. $\dagger$ indicates ensembles where analysis is still in progress.}  
\label{datatable}
	\end{table}

\section{Calculating $\mathbf{f_B}$ on the lattice}
\label{latticefB}
The decay constant $f_B$ is calculated from its lattice counterpart $\Phi_B$ via the equation
\begin{equation*}
f_B = \frac{1}{a} Z_\Phi \left[ \Phi_B^0 + c_A \Phi_B^1 \right]
\end{equation*}
where $\Phi_B$ is calculated from two-point correlators for axial and pseudoscalar operators:
\begin{equation*}
\Phi_B = - \frac{\sqrt{2 M_B} \mathcal{C}_{AP} }{\mathcal{C}_{PP}}\text{,} \qquad \qquad
\mathcal{C}_{AP} = \frac{\mel{\Omega}{A_4}{B} \mel{B}{P}{\Omega}}{2M_B}\text{,} \qquad \qquad \mathcal{C}_{PP} = \frac{\mel{\Omega}{P}{B} \mel{B}{P}{\Omega}}{2M_B}
\end{equation*}
and $Z_\Phi$ is calculated:
\begin{equation*}
Z_\Phi = \rho_A^{bl} \sqrt{Z_V^{bb} Z_V^{ll} }.
\end{equation*}
The perturbative constant $\rho_A^{bl}$ is set to 1 in this work, and similarly the higher-order correction coefficient $c_A$ in $f_B$ is set to 0. For determining $Z_V^{bb/ll}$, we compute meson three point functions of the vector current and enforce charge conservation. This formulation of $\Phi_B$ is equivalent to that used in \cite{fBWitzel}.

In these proceedings, we focus on the SU(3) flavour breaking effects in the ratio $f_B$/$f_{X_B}$ so that most sources of systematic error cancel. The SU(3) breaking in these ratios of $f_B$ or $f_{B_S}$ can be modelled by adapting the partially-quenched NLO equations in Bornyakov et al\cite{fRatio}:

\begin{equation}
\begin{split}
\label{fBequation}
\frac{f_B(q\overline{b})}{f_{X_B}} = &~1 + G(\delta \mu_q) + (H_1 + H_2)\delta \mu_q^2\\
&-(\tfrac{2}{3}H_1 + H_2)(\delta m_u^2 + \delta m_d^2 + \delta m_s^2) \\
&+ \ldots
\end{split}
\end{equation}
where $\delta \mu_q$ represent the distance between the valence quark mass and the SU(3) symmetric mass, and similarly $\delta m$ are differences in the sea quark masses. By fitting the coefficients in this equation to $f_{B}$ and $f_{B_s}$ on each set of ensembles, we can use the $\delta \mu$ and $\delta m$ values corresponding to the physical point to extrapolate to a physical prediction of $f_B$/$f_{X_B}$.

\section{Results}
\begin{figure}[htbp]
	\centering
	\includegraphics[width=0.85\textwidth]{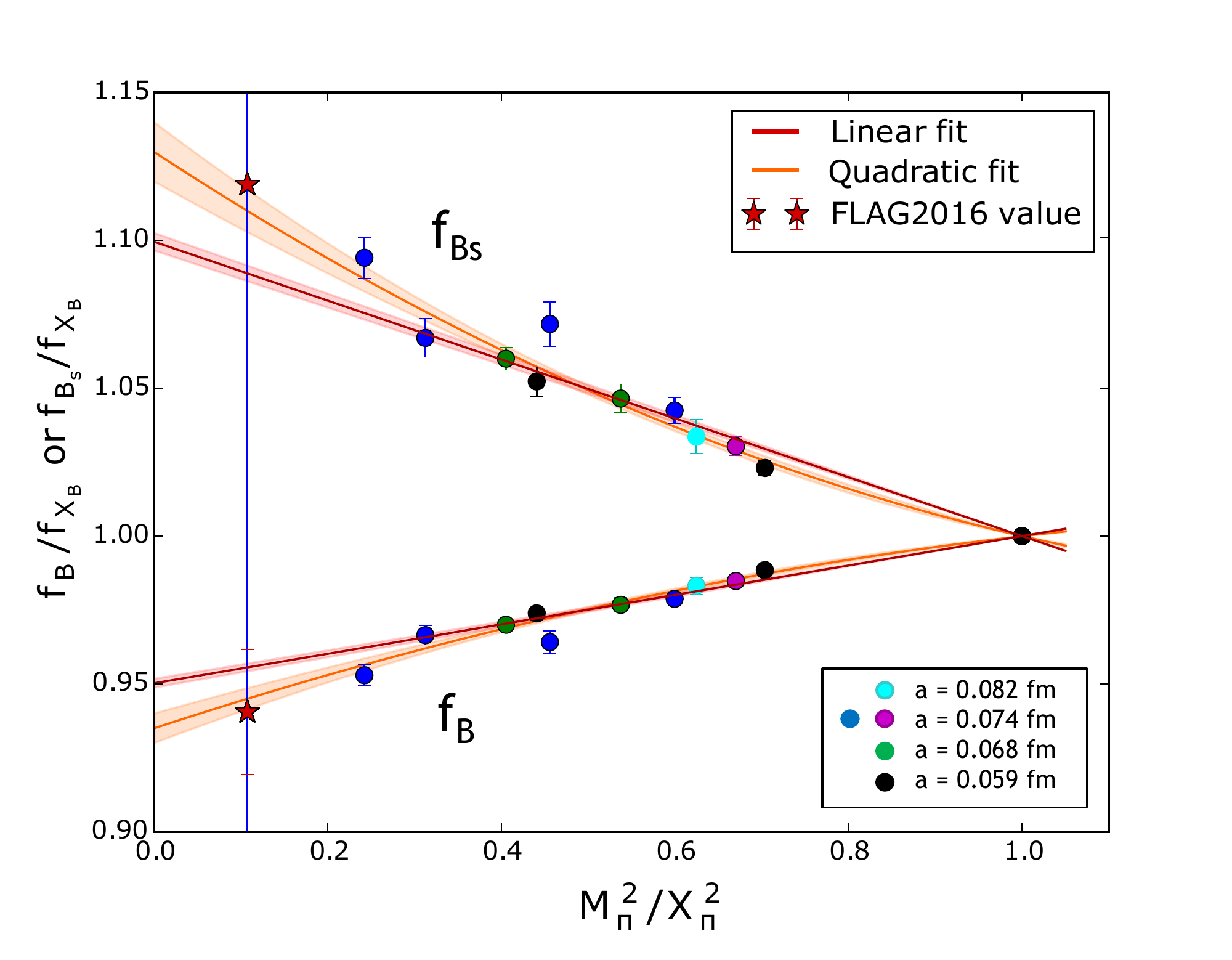}
	\caption{$f_B$/$f_{X_B}$ and $f_{B_s}$/$f_{X_B}$ for a variety of lattice ensembles.}
	\label{SU3pion}
	\end{figure}

The decay constant ratios $f_B$/$f_{X_B}$ and $f_{B_s}$/$f_{X_B}$ are shown against the SU(3) breaking ratio $m_\pi^2$/$X_\pi^2$ in Figure \ref{SU3pion}. It should be noted that the statistical error is mostly a result of error propagation resulting from interpolating the fit to the `best' $B$ meson on each ensemble. Both linear and quadratic fits in $m_\pi^2$ are shown: it can be seen that while the data is mostly linear in the breaking ratio $m_\pi^2$/$X_\pi^2$, an additional quadratic term is required to describe the data well and approach the FLAG values\cite{FLAG2016} denoted by $\star$.

\begin{figure}[htbp]
	\centering
	\includegraphics[width=0.82\textwidth]{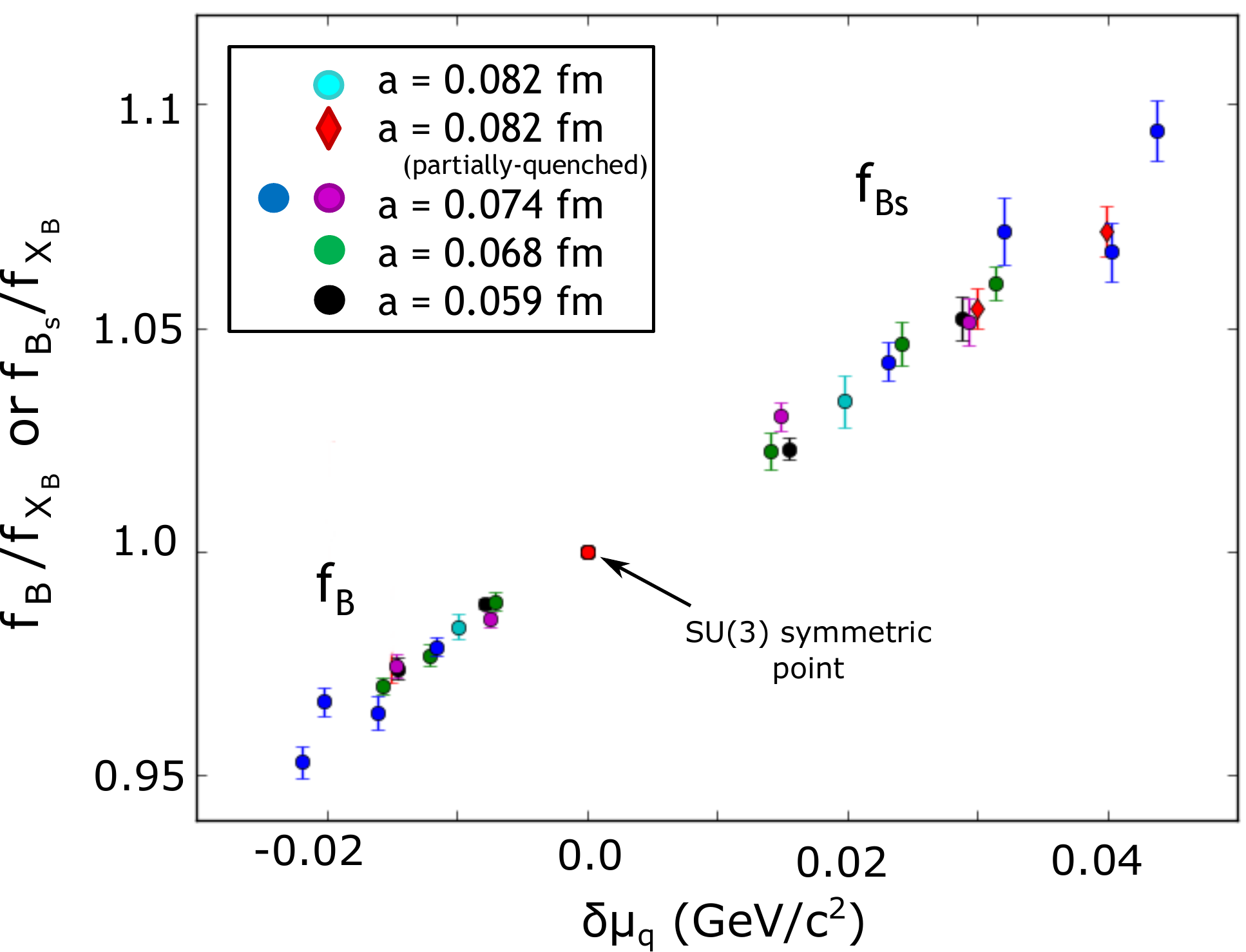}
	\caption{$f_{B_q}$/$f_{X_B}$ against $\delta \mu_q$ in physical units.}
	\label{SU3mq}
	\end{figure}

The same ratios are presented again in Figure \ref{SU3mq}, in terms of the light or strange quark mass $\delta \mu_q$ in Equation \ref{fBequation}. We also include some additional $B$ mesons with partially quenched light or strange quarks from the $32^3\times 64$ $\beta$=5.4 $\kappa_l = \kappa_s = 0.11993$ ensemble. Each set of ensembles with a different lattice spacing will require a separate fit as part of our future extrapolation toward the continuum limit. We note that the trend is mostly linear in $\delta\mu_q$, indicating that these terms in Equation \ref{fBequation} will dominate in agreement with observations made in the light quark sector\cite{fRatio}, although some hints of curvature (and thus higher order terms) are still visible for larger $\delta\mu_q$.

\section{Conclusion}
We have presented the current status of our investigation  of SU(3) symmetry breaking effects in $B$ meson decay constants. A large number of lattice ensembles have already been processed, though a few additional ensembles are required to complete our analysis of lattice discretisation effects and make extrapolations toward the physical and continuum limits in the near future. Studies of the $B^*$ decay constant are also in progress. 

\FloatBarrier
\acknowledgments
The numerical configuration generation (using the BQCD lattice
QCD program \cite{haar}) and data analysis
(using the Chroma software library \cite{Chroma}) was carried out
on the IBM BlueGene/Q and HP Tesseract using DIRAC 2 resources
(EPCC, Edinburgh, UK), the IBM BlueGene/Q at NIC (J\"ulich, Germany), the Cray XC40 at HLRN (The North-German Supercomputer
Alliance), the NCI National Facility in Canberra, Australia, 
and the iVEC facilities
at the Pawsey Supercomputing Centre. These Australian resources are provided
through the National Computational Merit Allocation Scheme and the
University of Adelaide Partner Share supported by the Australian
Government.
This work was supported in part through supercomputing resources
provided by the Phoenix HPC service at the University of Adelaide.
The BlueGene codes were optimised using Bagel~\cite{BAGEL}.
The Chroma software library~\cite{Chroma}, was used in the
data analysis.
This investigation has been supported by the Australian Research
Council under grants FT120100821, FT100100005, FT130100018 and DP140103067 (RDY and JMZ).

\bibliographystyle{JHEP}
\bibliography{LATTICE2018}


\end{document}